# From Dark Energy to Exolife: Improving the Digital Information Infrastructure for Astrophysics

**Michael J. Kurtz and Alberto Accomazzi**

NASA Astrophysics Data System
Center for Astrophysics | Harvard & Smithsonian
60 Garden St., Cambridge, MA 02138, USA
{mkurtz,aaccomazzi}@cfa.harvard.edu

**Thematic Areas**: Planetary Systems, Multi-Messenger Astronomy

## Abstract

Some of the most exciting and promising areas of Astronomy research today are found at the boundaries of the discipline: the search for Exoplanets and Multi-Messenger Astronomy. In order to achieve breakthroughs in these research fields over the next decade, innovation and expansion of the digital information infrastructure which supports this research is required.

Astronomy has been well-served by the existence of an open, distributed network of data centers and archives. However, institutional barriers and differing research cultures have prevented cross-disciplinary collaborations, creating fragmented knowledge and stove-piped research activities. This must change in order for the broader community of scientists to work together and solve our most ambitious decadal challenges.

Interdisciplinary inquiry is best supported by bringing researchers together at the information discovery level. In order to cross the traditional disciplinary silos we must allow scientists both to explore new ideas and to gain access to new data and knowledge. This is best enabled by providing discovery platforms which allow them to explore and connect different research threads in the literature, identify communities of experts, access and analyze the related published datasets, measurements and catalogs.

———

*"I have three recommendations for the 116th Congress to support the sciences and foster innovation: First, government needs to take a more active role in creating scientific infrastructure." -- Paul Romer (2018 Nobel Prize in Economics)*

# Introduction

Currently, two of astronomy's most vital research directions are explicitly intertwined with related yet separate research fields: the study of Exoplanets and Multi-Messenger Astronomy (MMA).  Researching exoplanets clearly requires a synthesis of expertise between Astrophysics and Planetary Science.  As the field matures, increased knowledge of and linkages with Geophysics and Earth Science will become necessary.  Should exolife be discovered, a far greater breadth of research will immediately be required for its understanding: not just data from new and more sophisticated instruments, but a synthesis of human knowledge greater than has ever been attempted about any research issue.  As such, it represents the biggest single interdisciplinary problem to face our generation.

Similarly, research in Dark Matter and Dark Energy requires knowledge at the intersection of Astronomy and Particle Physics. With the discovery of gravitational waves and neutrino detections, the field of so-called Multi-Messenger Astronomy (MMA) is now entering its golden age and is ripe for further discoveries which can help unlock the mysteries of the Universe.  In addition to a massive instrumentation infrastructure, MMA research requires expertise from various areas of Physics, Computer Science and Instrumentation.

Exolife and MMA research both present formidable challenges and thus require collaboration between experts working in different disciplines, using different instruments, data formats, workflows, and having different cultures.  Having these scientists work successfully together requires both human and technical infrastructure: effective communication among experts in different disciplines, interoperability across data archives, novel computational approaches, and AI-driven discovery capabilities.

In this whitepaper we argue that a robust information infrastructure is needed to enable the research breakthroughs required to meet these challenges. We suggest that the scientific literature be seen as the central organizing point used to navigate interdisciplinary research fields.  A shared discovery platform built on the literature allows communities of experts from different fields to cross the traditional information silos.  As interdisciplinary research develops, these fields become organically connected and discoverable through topics, citations, and co-readership.  Further connecting the literature with data products increases discoverability of both and allows for the data to be more accessible by non-experts.

## Research Infrastructure in Astronomy

Research infrastructure serves as the foundation for research. Organizations, structures, instruments, collections, machines, etc. enable science. The recent few decades have produced a new digital infrastructure for Astronomy research consisting of several organizations, which, while independent, work together to form an integrated whole.  While telescopes and their instruments tend to become obsolete on a timescale of a few decades, archives and information systems are better able to remain relevant and become durable, thanks to continuous incremental improvements in their capabilities and by incorporating new data and technologies.  However, digital libraries require long-term commitments of resources: their infrastructure must be regularly maintained and modernized, because research into all aspects of Astronomy depends on and is accelerated by it (Borgman et al. 2016).

The continued existence and development of Astronomy's network of data centers and archives is of substantial strategic importance to all of Astronomy's future developments. Simply maintaining the status quo is not a viable option. Over the next decade, primary sensors will produce more data than currently exist in Astronomy, presenting new challenges for organizations which archive and distribute it (e.g. MAST, Chandra, HEASARC, IPAC, ESO, ESA, and several yet to start). To make full use of our new resources, new methods will need to be developed to store, collate, and disseminate the new information. Besides the data archives themselves, this has important implications for data curation centers such as the CDS, CADC, NED, NExScI, and for the literature database, the NASA Astrophysics Data System (ADS).

The ADS is different from other Astronomy archives and data centers in that its primary focus is not data measurements (numbers), but concepts (words). As a nexus between scientists, the research literature, and the underlying data, ADS curates the connections between the high-level science results and the evidence which they are built upon, including observations, catalogs, measurements, and software. Thus ADS has a unique role in the practice of Astronomy, and it has unique opportunities and responsibilities both now and into the future.

The ADS has been serving astronomers for more than a quarter of a century, continuously improving in ways both seen and unseen. It is difficult to evaluate the impact of infrastructure: for instance, what impact does the Brooklyn Bridge have on New York City? Using a simple research time saved argument, Kurtz et al (2005) showed that the ADS's contribution to Astronomy research in 2003 was equivalent to the contribution of all the astronomers in France. Use of the ADS has increased substantially since then.

Libraries have always had the role of enhanced memory, a key component of intelligent human thought. The ADS is a library and its role is to collaborate with the scientist to perform research. As research and researchers continuously become more sophisticated, the ADS must also continue to implement and create new and improved methods and technologies in order to maintain its role as a trusted mechanical-intellectual partner for astronomers.

The need for machine assistance increases as the interdisciplinarity of research increases. Humans are not Darwinian evolving to be any smarter: rather, human intellectual evolution is driven primarily by machines (e.g. pens, the printing press), making our collective memories persistent and available. One might argue that there is perhaps no single individual who is both an expert on the spectroscopy of exoplanet atmospheres and on the theoretical aspects of earth tides. Therefore, an informed discussion of tides on exoplanets likely requires research level information from both realms. It is the role of the ADS to provide the means to discover this information, even by a scientist who may have no deep knowledge of either sub-field.

## The Search for Exolife

The discovery of exolife (if it happens at all) could well occur within the next decade. This would be as important a discovery as has ever happened, surely at least on the scale of the Copernican/Hubbelian discovery that the universe does not revolve around us. The current research into exoplanets already requires a much closer relationship between Astronomy, Planetary Science, Heliophysics and Geophysics. For the ADS to continue to provide and improve its service to this substantially enlarged community will require very substantial effort.

NASA is arguably the US agency which has developed the most sophisticated data infrastructure networks to support its science goals. Almost all of its data is open and accessible, which are necessary conditions to having it be FAIR: Findable, Accessible, Interoperable and Reproducible (Wilkinson et al, 2016). Making data findable, especially by cross-disciplinary researchers, requires an improved discovery ecosystem spanning disciplinary silos.

Despite the efforts made by NASA and other funders, thus far the research in the field of exoplanets has been conducted within the confines of each discipline. The number of papers studying exoplanets in core Astronomy journals has roughly doubled over the past 6 years and now makes up approximately 5% of the total refereed papers. For these papers, the influence of planetary journals has remained constant at approximately 9%, twice the average influence for all Astronomy articles (measured as the fraction of references to core Planetary Science journals). Over the same period, just under 1% of the papers in Planetary Science journals concerned exoplanets. This clearly shows that the two disciplines are still effectively disjointed, reflecting the stove-piped nature of research in the two fields, with most of the research currently being published in the main Astronomy journals.

The Exoplanet Science Strategy Report (NASEM 2018a), observes that "at NASA, research within the Science Mission Directorate is stove-piped into four divisions: Astrophysics, Planetary Science, Heliophysics, and Earth Science." The report further states that "the search for life outside the Solar System is a fundamentally interdisciplinary endeavor." Similarly, the Astrobiology Strategy Report (NASEM 2018b) observes that "integration across diverse and sometimes seemingly disparate disciplines is key to major progress on astrobiology's fundamental questions."

Studies of exolife will require widespread interdisciplinary collaboration. If some future exoplanet molecular spectroscopist notices certain anomalies in the Polycyclic Aromatic Hydrocarbon (PAH) signatures in the spectra, it will likely be more than our spectroscopist can do to notice that the strange PAH anomalies are quite similar to PAH profiles after being "processed" by mollusks. This is exactly the type of connection that an advanced discovery platform will be able to do. Without these expert system capabilities we might never discover Clam World.

As hard as it is to imagine an earth ruled by giant lizards, it is harder still to imagine what exolife might be like. This is exactly the point: for us humans to discover and understand other forms of life on distant planets requires the combined knowledge of us all. We will need to continuously merge and build upon the ideas and discoveries of many individuals, and most probably for an extended period of time, to be able to understand what we are seeing with our instruments. The continuous connecting, merging, and debating of ideas and evidence is the role of journals and information systems, such as the ADS, and the scientists who use them.

The increased reach of interdisciplinary research requires systems such as ADS to go beyond their silos in order to connect the relevant research objects. Currently, the ADS has no formal, ongoing collaboration with the Planetary Data System (PDS), or other planetary data sources, which would enable it to link data sets, measurements, or feature classifications with the journal articles where they appear. These capabilities have existed among the Astrophysics data centers and the ADS for more than two decades, and have allowed astronomers to seamlessly cross the boundaries between archives and the literature.

Although many planetary scientists use the ADS as a literature database which indexes nearly the entire refereed literature in Planetary Science, there is no equivalent service which links Planetary Data to its literature. Merging the two disciplines at the information system level will be a large international project, and for the ADS and PDS to establish a close working relationship would seem a logical and necessary first step.

**Multi-Messenger Astronomy Research**

Multi-Messenger Astronomy is an area in which Astronomy is merging into Physics. Gravitational wave Physics, Particle Astrophysics, Neutrino Astrophysics, Physical Cosmology, Dark Matter, Dark Energy are all major research areas in which the two disciplines both operate. High Energy Physics has long had a literature system similar to ADS, called INSPIRE, which is jointly operated and funded by CERN, DOE and DESY. INSPIRE, ADS, and arXiv have had a close, semi-formal, working relationship for more than ten years. Many scientists working in the overlapping fields use all three services routinely.

Historically the High Energy Physics (HEP) community has supported data collation services such as the Particle Data Group (PDG) which is in some ways similar to NED or SIMBAD; INSPIRE has links to these data sources. Contrary to Astronomy, HEP does not have a tradition of maintaining primary data archives, such as that for the NuStar data maintained at HEASARC, which both stores the data and creates links to the literature stored and indexed by ADS. NASA has been the leader in developing a robust data archiving and access system for the science it supports, but much of Multi-Messenger Astronomy is funded by DOE, NSF and other international agencies. The great success of the NASA archive system suggests that other agencies would do well to emulate that success, or, perhaps better, join the effort.

The existence of curated high-level data products linked to the literature in ADS and INSPIRE will greatly increase their discoverability, re-use, and overall scientific impact at a fraction of the cost of the original missions and experiments (White et al 2009, Rebull et al 2017).

## Recommendations

**ADS should continue to cover the literature in all research areas of Astronomy and Astrophysics, including Solar Physics, Planetary Science, Exoplanets, and Multi-Messenger Astronomy.** It should continue to enhance its capabilities as the amount of research literature indexed in its system increases in volume and complexity and as scientists from multiple disciplines start using it as a discovery platform.

**NASA should encourage greater integration of Astrophysics and Planetary Science research, particularly as it affects Exoplanets.** This can begin with a closer collaboration between ADS and PDS, which would immediately increase the discoverability of planetary research topics and datasets, extending the connections between Astronomy and Planetary Science data through shared literature.

**DOE and NSF should emulate NASA in establishing, enhancing, and curating high-level data archives for their research portfolios, particularly as it affects MMA research.** The creation and development of well-curated repositories and organizations supporting MMA will provide the shared digital infrastructure required for this interdisciplinary research endeavor.